\documentclass[manuscript]{aastex}
\usepackage{epsfig}
\usepackage{color}
\shorttitle{HCN and SFR in M51}
\shortauthors{Chen et al.}

\begin{document}

%% LaTeX will automatically break titles if they run longer than
%% one line. However, you may use \\ to force a line break if
%% you desire.

\title{Spatially-Resolved Dense Molecular Gas and \\ Star Formation Rate in M51}

\author{Hao Chen\altaffilmark{1, 2, 3},}

\author{Yu Gao\altaffilmark{4, 5},}

\author{Jonathan Braine\altaffilmark{6, 7},}

\and

\author{Qiusheng Gu\altaffilmark{1, 2, 3}}

\email{chenhao198701@yahoo.com, yugao@pmo.ac.cn, braine@obs.u-bordeaux1.fr}

\altaffiltext{1}{School of Astronomy and Space Science, Nanjing University, Nanjing 210093, P. R. China}
\altaffiltext{2}{Key Laboratory of Modern Astronomy and Astrophysics, Nanjing University, Nanjing 210093, P. R. China}
\altaffiltext{3}{Collaborative Innovation Center of Modern Astronomy and Space Exploration， Nanjing, 210093, P.R. China}
\altaffiltext{4}{Purple Mountain Observatory, Chinese Academy of Sciences, 2 West Beijing Road, Nanjing 210008, P. R. China}
\altaffiltext{5}{Key Laboratory of Radio Astronomy, Chinese Academy of Sciences, Nanjing 210008, P. R. China}
\altaffiltext{6}{Univ. Bordeaux, Laboratoire d'Astrophysique de Bordeaux, F-33270, Floirac, France.}
\altaffiltext{7}{CNRS, LAB, UMR 5804, F-33270, Floirac, France}

\begin{abstract}
We present the spatially-resolved observations of HCN J = 1 -- 0 emission in the nearby spiral galaxy M51 using the IRAM 30 m telescope. 
The HCN map covers an extent of $4\arcmin\times5\arcmin$ with spatial resolution of $28\arcsec$, which is, so far, the largest in M51. 
There is a correlation between infrared emission (star formation rate indicator) and HCN (1--0) emission (dense gas tracer) at kpc scale in M51, a natural extension of the proportionality between the star formation rate (SFR) and the dense gas mass established globally in galaxies.
Within M51, the relation appears to be sub-linear (with a slope of 0.74$\pm$0.16) as $L_{\rm IR}$ rises less quickly than $L_{\rm HCN}$. 
We attribute this to a difference between center and outer disk such that the central regions have stronger HCN (1--0) emission per unit star formation.
The IR-HCN correlation in M51 is further compared with global one from Milky Way to high-z galaxies and bridges the gap between giant molecular clouds (GMCs) and galaxies.  
Like the centers of nearby galaxies, the $L_{\rm IR}$/$L_{\rm HCN}$ ratio measured in M51 (particularly in the central regions), is slightly lower than what is measured globally in galaxies, yet is still within the scatter.
This implies that though the $L_{\rm IR}$/$L_{\rm HCN}$ ratio varies as a function of physical environment in the different positions of M51, IR and HCN indeed show a linear correlation over 10 orders of magnitude.

\end{abstract}


\section{ INTRODUCTION }

The molecular gas in galaxies is organized in giant molecular clouds (GMCs).  Within these clouds, denser regions, often called clumps and cores, are found and that is where the star formation takes place.
The star formation rate (SFR)  is linearly correlated with the amount of molecular 
gas \citep{Kennicutt1989ApJ...344..685K, Bigiel2008AJ....136.2846B} with the exception of the infrared luminous galaxies, whose SFR is higher than would be expected from the CO emission (molecular gas tracer).  
The SFR and total gas (H$_2$ and HI) surface densities are correlated \citep{Kennicutt1998ApJ...498..541K}, typically with power-law exponents greater than one when the HI dominates, i.e. the lower column density regions generally in the outer disk, and approaching 1 in the H$_2$ dominated regions \citep{Buat1989A&A...223...42B, Buat1992A&A...264..444B, Bigiel2008AJ....136.2846B}. 
Thus, at the whole-galaxy scale, the important link is between the SFR and molecular gas content, particularly the mass of dense gas.
The star formation efficiency (SFE; SFR/CO) is higher in galaxies with a higher dense gas fraction \citep{Gao2004ApJ...606..271G}.
\citet{Lada2012ApJ...745..190L} suggest that the SFR and molecular gas correlation is always linear with the same dense gas fraction from clouds to galaxies.

A few environments exist in which the dense gas fraction, typically the HCN/CO intensity ratio, is clearly different from the average value of normal galaxies \citep[$\sim1/30$,][]{Gao2004ApJ...606..271G}.  
Ultraluminous infrared galaxies have a higher dense gas fraction, while environments having undergone a gas-gas collision typically have a lower dense gas fraction \citep[e.g.][]{Gao2004ApJ...606..271G, Braine2003A&A...408L..13B, Braine2004A&A...418..419B}.
   
The physical conditions of active star-forming GMC cores are revealed by emission from high-J ($>$3) rotational transition CO lines and large dipole-moment molecules like CS, HCO$^+$, and HCN \citep{Evans1999ARA&A..37..311E}.
The HCN $J=1-0$ transition traces denser gas ($\geq 3\times10^4\ cm^{-3}$) than CO $J=1-0$ ($\geq 300\ cm^{-3}$) because of its larger dipole moment.
From a HCN emission survey of  65 galaxies, \citet{Gao2004ApJ...606..271G} find a tight linear correlation between the SFR, traced by infrared (IR) luminosity ($L_{\rm IR}$), and dense gas mass ($M_{\rm dense}$), traced by the HCN luminosity ($L_{\rm HCN}$),  with an almost constant average ratio $L_{\rm IR}/L_{\rm HCN} = 900L_\odot/(K\ km\ s^{-1}\ pc^2)$, whereas the IR-CO correlation is clearly non-linear.
\citet{Wu2005ApJ...635L.173W, Wu2010ApJS..188..313W} extended this correlation to Galactic GMCs in the Milky Way. 
The HCN--IR correlation appears to be valid at high redshift as well \citep{Gao2007ApJ...660L..93G}, but a large gap is present between global galaxies and Galactic GMCs \citep[cf. Figure 2 in][]{Lada2012ApJ...745..190L}.

Few HCN maps of external galaxies are available \citep[e.g.,][]{Nguyen1992ApJ...399..521N, Gao2004ApJS..152...63G, Kepley2014ApJ...780L..13K} because the HCN emission in normal spiral disks is weak, such that little is known about the dense gas distribution outside the central regions in external galaxies. Except for a few high-resolution HCN maps \citep{Kohno1996ApJ...461L..29K, Helfer1997ApJ...478..162H, Schinnerer2007A&A...462L..27S, Krips2007A&A...468L..63K, Levine2008ApJ...673..183L, Muraoka2009PASJ...61..163M, Leroy 2015ApJ...801...25L} of the central kpc and a few single-dish maps of IC 342 ($40\arcsec\times60\arcsec$) and  M83 ($40\arcsec\times50\arcsec$) and M51 \citep[$1\arcmin.5\times2\arcmin$,][]{Nguyen1992ApJ...399..521N}, few HCN observations exist toward the outer disks.  
More observations of HCN emission in nearby galactic disks are necessary in order to properly interpret whole-galaxy data which is a mixture of nucleus and disk emission.

M51 is an ideal target to study the spatially resolved IR-HCN correlation across the galactic disks. It is a large local  spiral galaxy \citep[D $\sim$ 7.6 Mpc,][]{Ciardullo2002ApJ...577...31C} seen nearly face-on \citep[i $\sim 22^\circ$,][]{Colombo2014ApJ...784....4C}. 
The Herschel\footnote{Herschel is an ESA space observatory with science instruments provided by European-led Principal Investigator consortia and with important participation from NASA.} Very Nearby Galaxy Survey (VNGS) mapped M51 at wavelengths of 70, 160, 250 and 350 $\micron$ \citep{Mentuch2012ApJ...755..165M}. CO emission has been mapped over the entire galaxy by \citet[CO 1--0]{Koda2011ApJS..193...19K}, \citet[CO 2--1]{Schuster2007A&A...461..143S} and \citet[CO 3--2]{Vlahakis2013MNRAS.433.1837V}. 
The PdBI Arcsecond Whirlpool  Survey \citep[PAWS,][]{Schinnerer2013ApJ...779...42S} observed the CO 1--0 emission in the central 9 kpc of M51 at 1$\arcsec$ resolution.  
HCN emission in the central disk ($1\arcmin.5\times2\arcmin$) has been mapped by \citet{Nguyen1992ApJ...399..521N}. 
The correlations between SFR and H$_2$ at kpc scale in M51 are explored by \citet{Kennicutt2007ApJ...671..333K} and \citet{Bigiel2008AJ....136.2846B}. 

In this paper, we present a map of M51 in the HCN $J = 1 - 0$ line at 88.6 GHz with the IRAM 30 meter telescope to obtain a spatially resolved HCN image of a nearby spiral galaxy covering $4\arcmin\times5\arcmin$.
The IRAM 30m beam corresponds to a spatial resolution of $\sim$1 kiloparsec and samples large GMC associations and individual GMCs within the beam along the spiral arms to starburst-like regions near the center. Combined with the Herschel VNGS 70 $\micron$ and 160 $\micron$ maps, we can make a detailed point-by-point comparison of the IR-HCN correlation at kpc scales within a normal star-forming disk, bridging the luminosity gap between galaxies and individual GMCs. 

The Observations and data reduction are presented in Section 2. 
The HCN maps and the kpc-scale IR--HCN correlation are presented in Section 3. 
In Section 4, we compare the distribution of HCN, CO and IR, and discuss the effect of dust temperature on the HCN emission.  
The IR--HCN correlation in M51 is compared with existing data for galaxies and Galactic clouds. 

\section{OBSERVATIONS AND DATA REDUCTION}

\subsection{HCN Observations}

During May 2004, June 2005 and July 2007, we used the IRAM 30 meter telescope on Pico Veleta near Granada, Spain, to map the HCN $J = 1 - 0$ emission, at a rest frequency $\nu = 88.63185$ GHz, from M51. The angular resolution of the IRAM 30m is $28\arcsec$ at 88.6 GHz, which is $\sim$1 kpc at the distance of M51.
The map covers $4\arcmin\times5\arcmin$  (8 kpc $\times$ 10 kpc) and the spectra are spaced by $15\arcsec$ (see Figure 1).  So far, this is the largest HCN map of M51.

Pointing were checked about every 2 hours on nearby QSOs with strong millimeter continuum emission or occasionally planets, which provided a pointing accuracy of better than $3\arcsec$.
The A100 and B100 receivers which receive two orthogonal polarizations were used to observe the molecular lines simultaneously. 
We used the 512 $\times$ 1MHz filter bank backend, yielding a spectral resolution of 3.3 km/s.
Typical system temperatures at 88.6 GHz were $\sim$130 K on average on the $T_A^*$ scale.
To convert the observed antenna temperatures to the main-beam temperature scale, we use $T_{\rm mb}$ = $T_{\rm A^*}$ $F_{\rm eff}/B_{\rm eff}$ where the forward efficiency $F_{\rm eff}$ = 95\% and the beam efficiency $B_{\rm eff}$ = 78\%.
Observations focused on the spiral arms for the outer disk regions due to the limited observing time available.
Some bad weather data were discarded and the total observing time of the useful data in M51 is $\sim$ 81 hours.
The integration time at each position was 6 -- 90 minutes depending on the signal strength.

All the data were reduced with the CLASS program of the GILDAS package\footnote{http://www.iram.fr/IRAMFR/GILDAS/}. 
After eliminating bad channels, 
a linear baseline is fit to correct the total power variations in the receiver and atmosphere.  A small fraction of the spectra are not well fitted by a linear baseline.
The line windows are determined from the HCN spectra themselves when the emission is strong.  Where the emission is weak or not detected, we use line windows based on the CO emission, which is detected throughout the map.
Individual spectra with poor baselines are discarded and spectra at each position are combined.

\subsection{CO Data} 

The CO J = 1 - 0 total power and short spacing data were taken with the $5\times5$ Beam Array Receiver System on the Nobeyama Radio Observatory 45 m telescope \citep{Koda2011ApJS..193...19K}.

All CO and the Herschel IR data (70, 160 and 250 \micron) are convolved to the resolution of the HCN observations ($28\arcsec$) to make a detailed point-by-point comparison with HCN. 

\section{RESULTS AND ANALYSIS}

\subsection{HCN line}

\subsubsection{HCN Spectra}

\begin{figure}
\centering\includegraphics[angle=0,scale=.9]{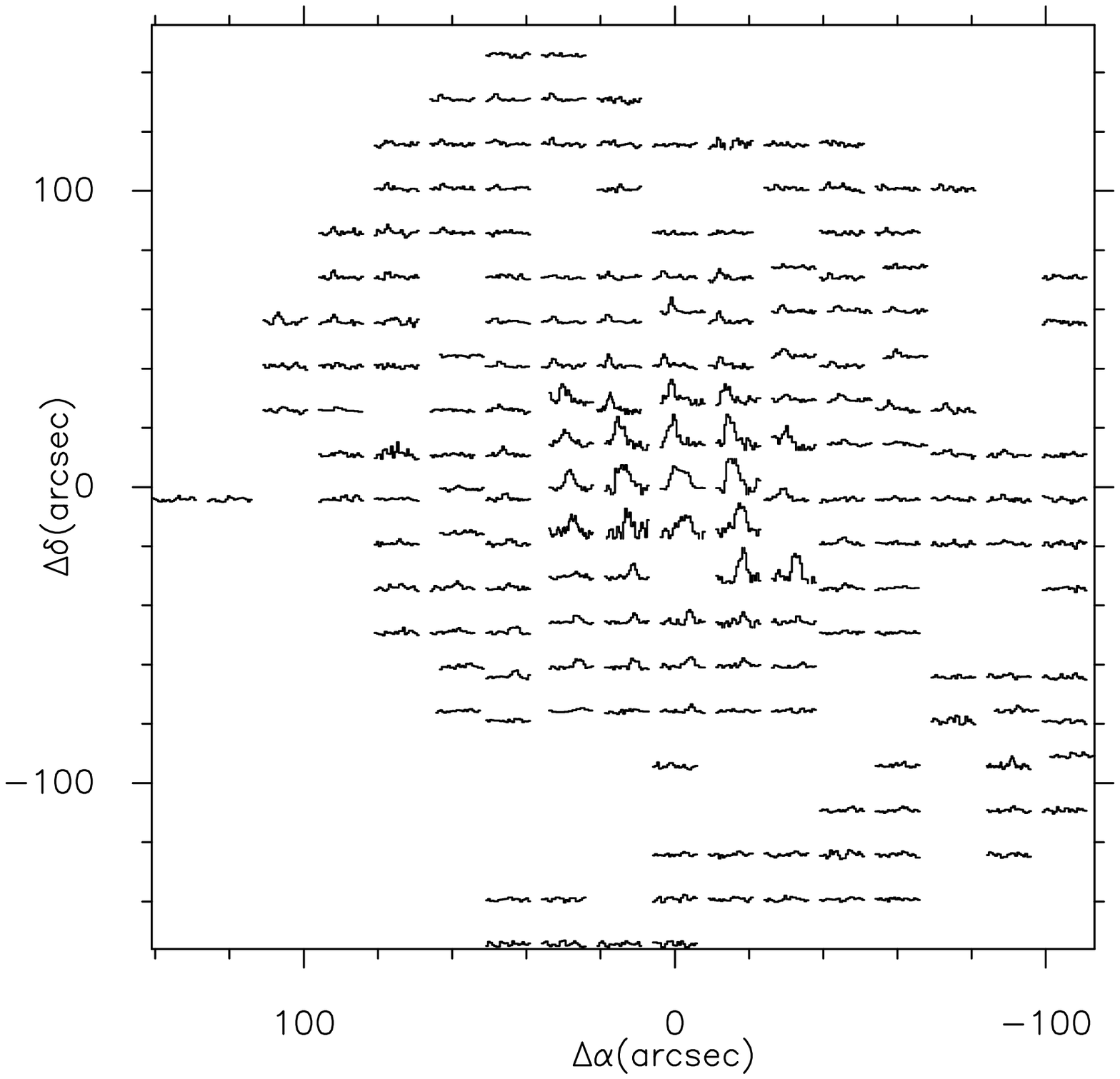}
\caption{The HCN $J = 1 - 0$ line profiles observed in M51. Each profile is represented as a function of the antenna pointing offset in right ascension (RA) and declination (DEC) relative to the map center ($\alpha$ = $13^h29^m52.7^s$, $\delta$ = $47^{\circ}11\arcmin43.0\arcsec$). The vertical and horizontal axes of each spectrum
are $T_{\rm mb}$ (K) from -0.01 to 0.06 mK and $V$ ($km\ s^{-1}$) from 320 to 620 $km\ s^{-1}$, respectively. All the data have been smoothed to a velocity resolution of 13.3 $km\ s^{-1}$ for show (for some weak HCN positions, the spectra are further binned to 26.7 $km\ s^{-1}$ for display).}
\end{figure}

\begin{figure}
\centering\includegraphics[angle=0,scale=.58]{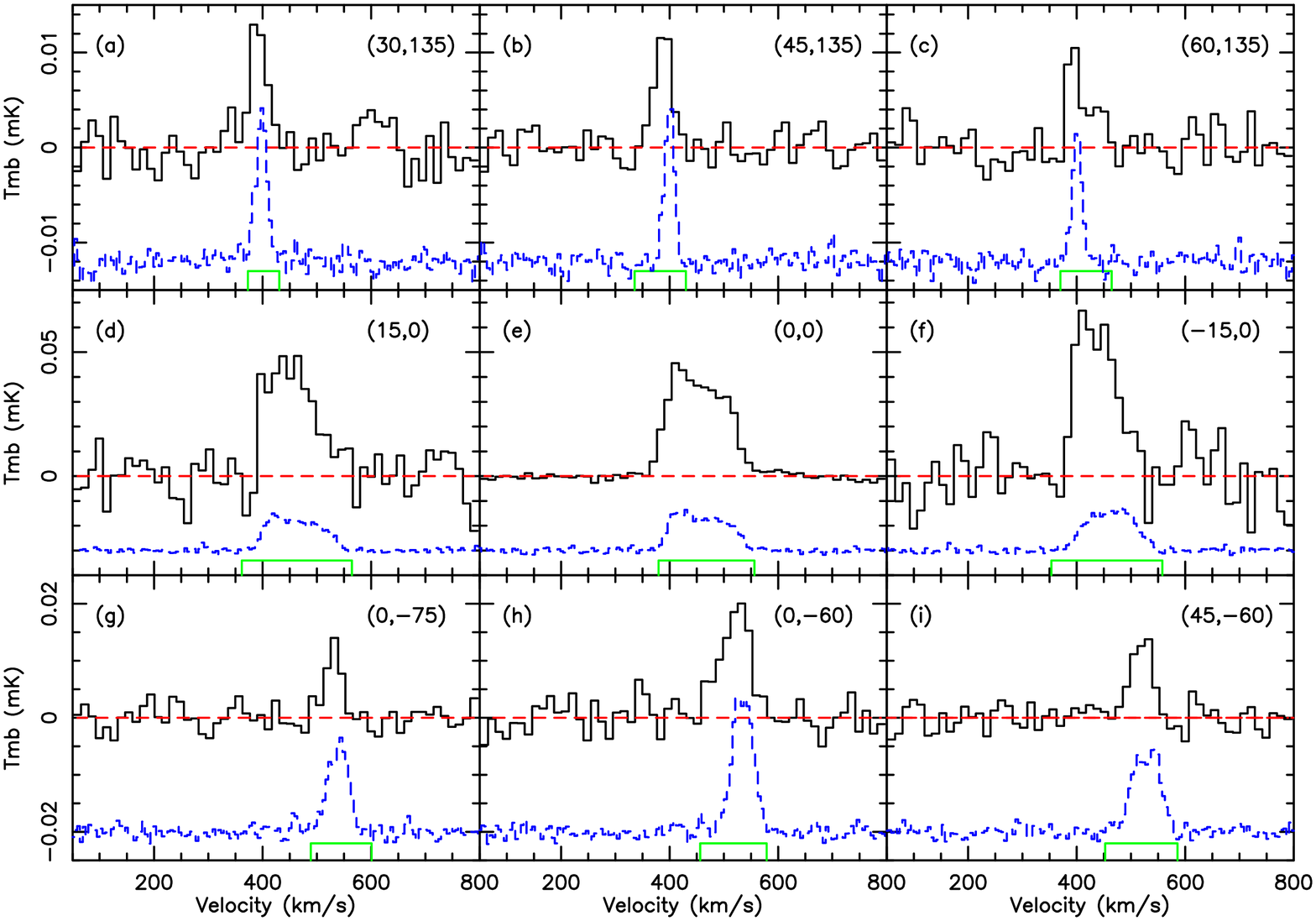}
\caption{Spectra of HCN (black solid lines) and CO (blue dashed lines) at representative positions in M51. All spectra are on the $T_{\rm mb}$ scale and binned to a velocity resolution of 6.7 $km\ s^{-1}$ for HCN and 5.2 $km\ s^{-1}$ for CO. 
The temperature scale of CO spectra are divided by 30 
for comparison purposes. Coordinates of the observed positions ($\Delta\alpha,\Delta\delta$) are indicated at the top right and the HCN integrated line windows are shown at the bottom in each box. }
\end{figure}

The HCN $J = 1 - 0$ spectra are shown in Figure 1.
All spectra are plotted in main beam temperature ($T_{\rm mb}$) and smoothed to 13.3 $km\ s^{-1}$ resolution for show (for some weak HCN positions, the spectra are further binned to 26.7 $km\ s^{-1}$ for display). 
Figure 2 shows several HCN and CO line profiles in detail. The HCN emission line is centered at 475 $km\ s^{-1}$ in the nucleus and shifted to 535 and 400 $km\ s^{-1}$ in the northern and southern regions respectively. 
The peak of HCN spectrum is 1/10 of the CO peak in the center regions while it is 1/30 in the outer disk.
The emission windows of the HCN lines are very similar to that of CO lines and the line widths tend to decrease as the galactocentric distance increase. In the central regions, the zero power line width is close to 200 $km\ s^{-1}$ and decreases to 30 -- 60 $km\ s^{-1}$ in the outer disk.
The HCN line peak and profile in the center are consistent with the previous observations of \citet{Nguyen1992ApJ...399..521N}.

\subsubsection{Local HCN Line Luminosities}

The HCN integrated intensities were measured as $I_{\rm HCN}$ = $\int T_{\rm mb}$$d$V over the emission window. 
For the HCN spectra with high signal to noise ratio ($S/N\geq5$) in integrated intensity, we define the emission window from HCN. But for the HCN spectra with low intensity S/N ($<5$), we define the emission window using the CO spectra, detected in all positions.
As the intensity is depend on the strength of line relative to zero point defined by the baseline outside the line window  \citep{Braine1993A&AS...97..887B, Matthews2001ApJ...549L.191M}, the integrated intensity errors were estimated as
\begin{center}
$\delta_{\rm HCN} =  T_{\rm rms}\sqrt{W_{HCN}\delta_c/(1-W_{\rm HCN}/W)},$ \\
\end{center}
where $T_{\rm rms}$ is the rms noise fluctuation, $W_{\rm HCN}$ is the width of the line window in $km\ s^{-1}$, $\delta_{\rm c}$ is the channel spacing in units of $km\ s^{-1}$, and W is the entire velocity coverage (1000 $km\ s^{-1}$). 3$\delta_{\rm HCN}$ upper limits were established for the weak $S/N<3$ spectra.

\begin{figure}
\centering\includegraphics[angle=0,scale=0.58]{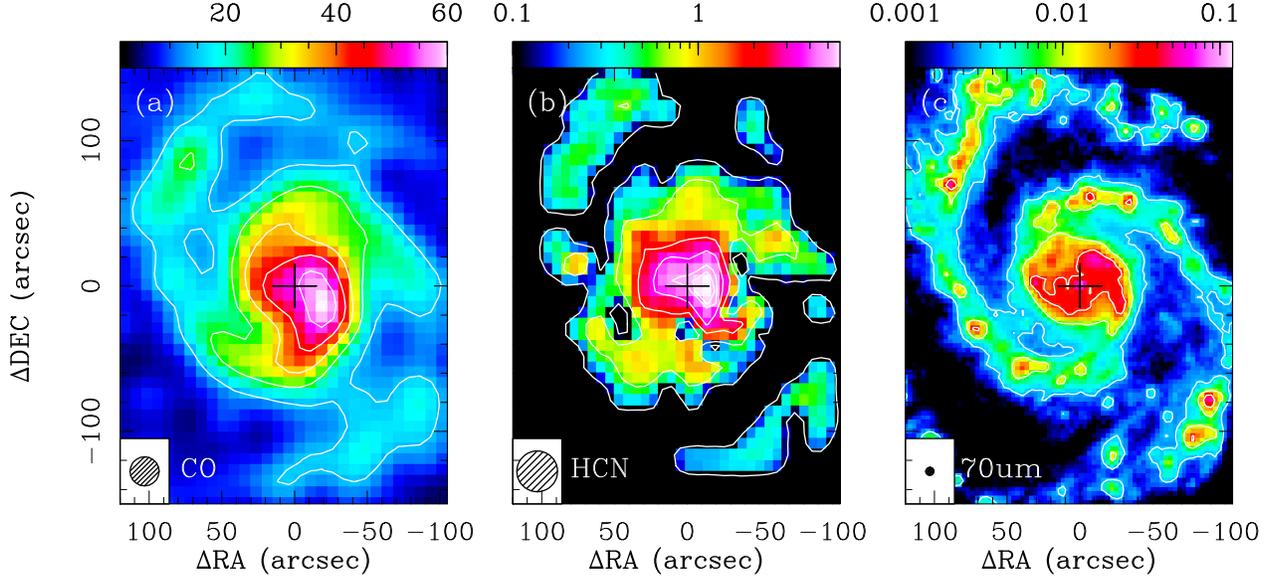}
\caption{(a) Map of CO 1--0 integrated intensity $[K\ km\ s^{-1}]$ showing M51 \citep{Koda2011ApJS..193...19K}; contours: $I_{\rm CO}$ = 14.7 24.7 34.7 44.7 54.7 $K\ km\ s^{-1}$; top: color scale in $K\ km\ s^{-1}$. (b) Map of HCN 1--0 integrated intensity  $[K\ km\ s^{-1}]$; contours: $I_{\rm HCN}$ = 0.1, 0.6, 1.9, 3.4, 4.9, 5.4 $K\ km\ s^{-1}$; top: color scale in $K\ km\ s^{-1}$. (C) Map of 70 $\micron$ intensity [Jy/pixel] (with a pixel size of $1.4 \arcsec$); contours: $I_{\rm 70 \micron}$ = 0.003, 0.009, 0.027, 0.081 Jy/pixel; top: color scale in Jy/pixel. The beam size is shown in the lower left corner. 
The cross marks the [0,0] center position at $\alpha$ = $13^h29^m52.7^s$, $\delta$ = $47^{\circ}11\arcmin43.0\arcsec$ (J2000).
All the antennae temperatures involved in the integrated intensities are on the $T_{\rm mb}$ scale.}
\end{figure}

The HCN luminosity in each observed region is calculated as:
\begin{center}
$L_{\rm HCN} (K\ km\ s^{-1}\ pc^2)= 23.5\Omega d_L^2(Mpc)I_{\rm HCN}(K\ km\ s^{-1}),$
\end{center}
where $\Omega$ is the solid angle of the regions covered by the telescope beams in $arcsec^2$ \citep{Solomon1997ApJ...478..144S} and $d_L$ is the luminosity distance. The local CO luminosity is calculated with the same formula (switching HCN with CO). 

\subsubsection{HCN Distribution}

Figure 3 shows the CO, HCN and 70 $\micron$ integrated intensity maps respectively. 
The HCN integrated intensity peak is 5.54$\pm$0.04 $K\ km\ s^{-1}$, located at 15$\arcsec$ west of the optical nucleus and about 15$\arcsec$ north of the CO peak (relative to the HCN resolution).  
The HCN map reproduces the large-scale structure traced by CO fairly well. The northern and southern CO arms are detected in HCN.
70 $\micron$ image (traces SFR) shows the same structure with HCN and CO. The high resolution 70 $\micron$ peaks in spiral arms are tightly relate to HCN peaks.

\begin{figure}
\centering\includegraphics[angle=0,scale=.95]{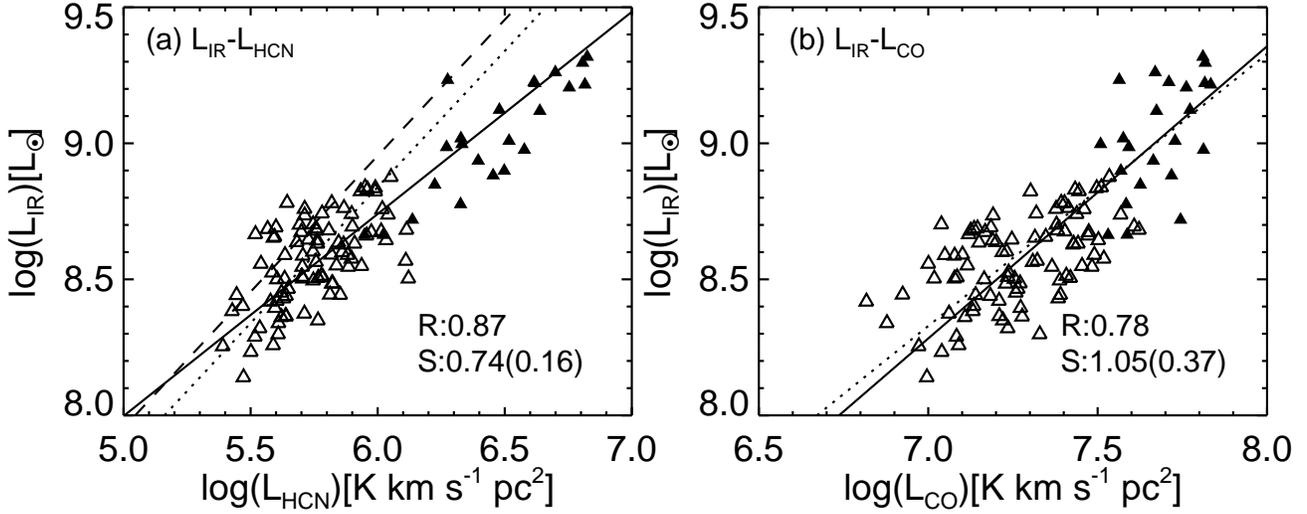}
\caption{IR (y-axis) as a function of HCN 1--0 and CO 1--0 (x-axis). Each point corresponds to an individual pointing. 
The central regions (filled triangles) and outer disk (open triangles) regions are distinguished in these correlations.
The solid lines show the best fit. The dashed line shows consistent  $L_{\rm IR}/L_{\rm HCN}$ value (900 $K\ km\ s^{-1}\ pc^2$) derived from the global galaxies \citep{Gao2004ApJ...606..271G}. 
Dotted lines indicate average ratio of IR/HCN in the outer disk.}
\end{figure}

\subsection{Correlation between HCN and IR}

\subsubsection{ $L_{\rm IR}$ from 70 $\micron$ and 160 $\micron$ fluxes}

In order to compare the IR and HCN point by point, we need to calculate the IR luminosity of each HCN beams.  The Herschel PACS observations at 70 and 160 $\mu$m bracket the peak of the thermal dust emission and as such provide an excellent estimate of the dust luminosity \citep{Boquien2011A&A...533A..19B}. 
Following \citet{Galametz2013MNRAS.431.1956G}, we calculate the total IR luminosity from the convolved 70 $\micron$  and 160 $\micron$ images as:
\begin{center}
$\log L_{\rm IR} = [(0.973\times \log \nu L_{\nu}(70) + 0.567) + (1.024\times \log \nu L_{\nu}(160)+ 0.176)]/2.$
\end{center}

\subsubsection{Correlation between HCN and IR}

Figure 4 shows $L_{\rm IR}$ vs. $L_{\rm HCN}$ for all of the points with HCN detection ($S/N \geq 3$).
The HCN emission is correlated tightly with the IR over 1.5 orders of magnitude in HCN intensities (correlation coefficient R = 0.87). The least-square fits yield a slope of 0.74, such that the SFR varies less than the dense gas mass. The best-fit (logarithmic) relation is:
\begin{equation}
\log(L_{\rm IR}(L_{\odot})) = (0.74\pm0.16)\log(L_{\rm HCN}(K\ km\ s^{-1}\ pc^2)) + (4.28\pm0.95).
\end{equation}
The uncertainties given in the equation refer to the fitting errors responding to the uncertainty of $L_{\rm HCN}$. 
The dense gas star formation efficiency ($L_{\rm IR}/L_{\rm HCN}$) of the central regions (filled triangles, with a average $L_{\rm IR}/L_{\rm HCN}$ ratio of 388 $\pm$ 47 $L_{\odot}(K\ km\ s^{-1}\ pc^2)^{-1}$, just the HCN luminosity errors are taken in account) is systematically lower than that of the outer disk regions (open triangles, with a averaged  $L_{\rm IR}/L_{\rm HCN}$ ratio of 691 $\pm$ 156 $L_{\odot}(K\ km\ s^{-1}\ pc^2)^{-1}$).

\subsubsection{Comparison of the HCN-IR and CO-IR correlations}

The relation between CO and IR luminosity of the same regions in M51 is also shown in Figure 4. The least-square fit yields a slope of 1.07$\pm$0.28 (correlation coefficient R = 0.78), such that the HCN varies more than the IR and CO for the same regions. 
The correlation coefficient between HCN and IR (0.87) is better than that of the CO-IR correlation (0.78).
The fitting error for the slope of IR-CO (0.28) is much larger than that of IR-HCN (0.16).  
The most obvious difference occurs at the high $L_{\rm IR}$ end, dominated by the central region, similar to the situation for galaxies as a whole. 
In Figure 4, lines of constant IR/HCN and IR/CO ratios fit the outer disk regions, but the constant IR/HCN line lies above most of the central regions, while the constant IR/CO line does not. 

\section{DISCUSSION}

\begin{figure}
\centering\includegraphics[angle=0,scale=.8]{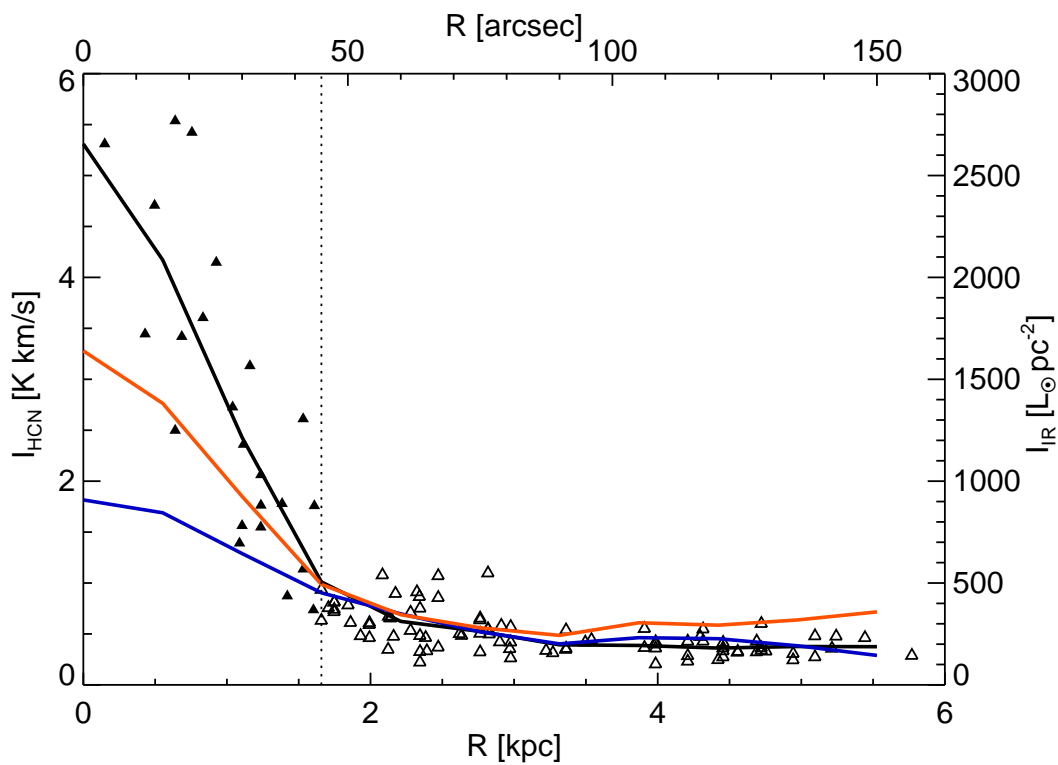}
\caption{Intensity of HCN in the central regions (filled triangles) and in the outer disk (open triangles) as a function of the offset from the assumed center of M51 ($\alpha$ = $13^h29^m52.7^s$, $\delta$ = $47^{\circ}11\arcmin43.0\arcsec$). 
Average values of CO (blue, divided by 30),  HCN (black) and IR (red) within the intervals $r=n\times15\arcsec+7.5\arcsec\pm7.5\arcsec$ where n = 0, 1...10 are displayed by lines. Dotted line show the radii of 45\arcsec (1.66 kpc) to distinguish the center and outer disk.}
\end{figure}

\begin{figure}
\centering\includegraphics[angle=0,scale=.8]{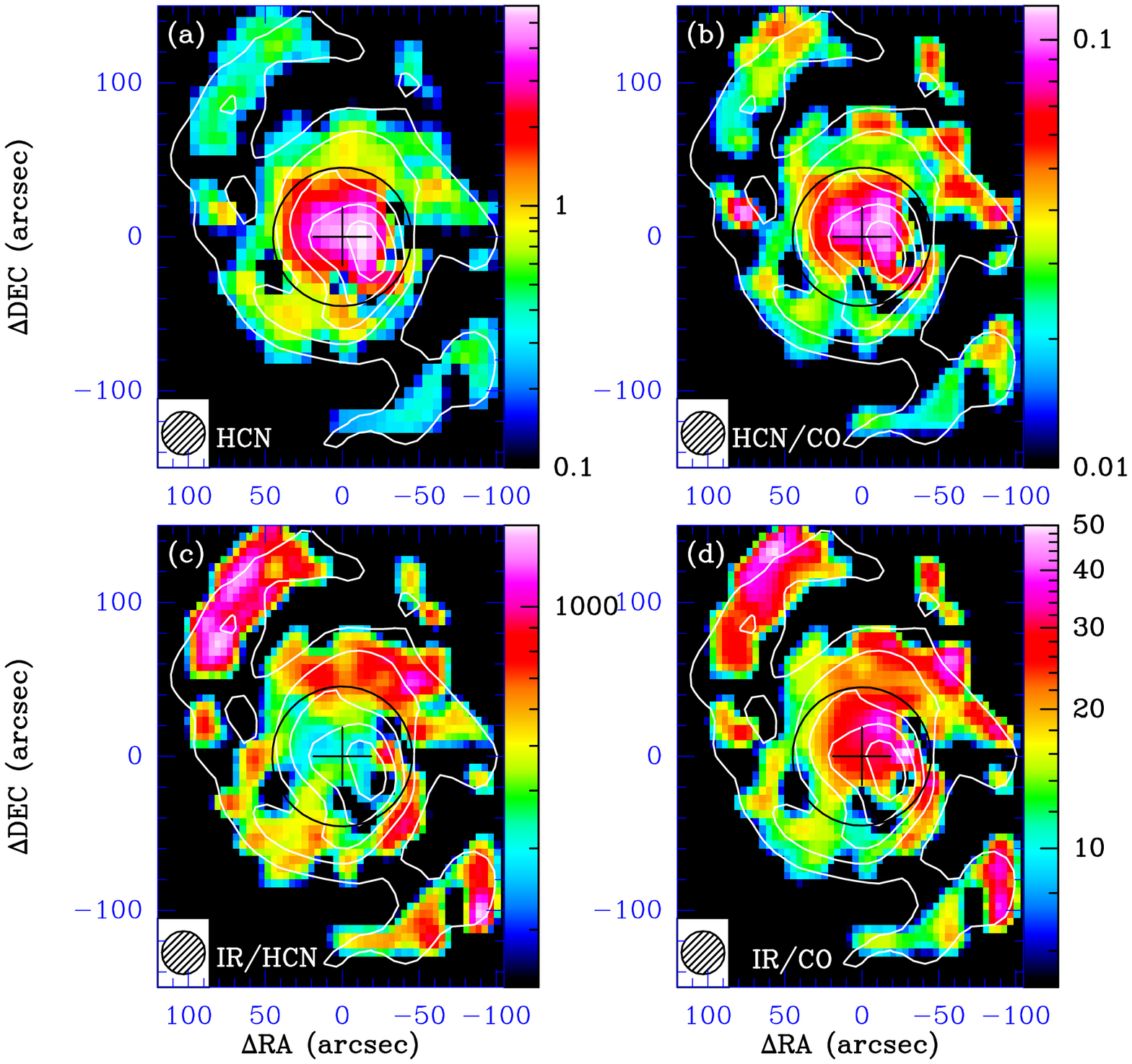}
\caption{(a) HCN J = 1 - 0 integrated intensity map $[K\ km\ s^{-1}]$ of the detected regions ($S/N\geq3$).
(b) (c) (d) are respectively maps of $L_{\rm HCN}$/$L_{\rm CO}$, $L_{\rm IR}$/$L_{\rm HCN}$ $[L_\odot/(K\ km\ s^{-1}\ pc^2)]$ and $L_{\rm IR}$/$L_{\rm CO}$ $[L_\odot/(K\ km\ s^{-1}\ pc^2)]$ of the same region. Contours are the same as those in Figure 3(a). The black circle shows the assumed center occupying the central 45\arcsec (1.66 kpc).}
\end{figure}

\subsection{Comparison of the HCN, CO and IR Distribution}

The azimuthal averages are calculated to investigate the radius distribution of CO, HCN and IR displayed as lines in Figure 5. 
We average over rings of $15\arcsec$ width (regions within the intervals $r = n\times15\arcsec+7\arcsec.5\pm7\arcsec.5, n = 0, 1...10)$, comparable to the angular resolution of our data. 
All the intensities of CO, HCN and IR decrease rapidly with radius in the central 45\arcsec (1.66 kpc) and then more slowly at larger radii.
HCN decreases faster than IR, and IR decreases faster than CO along with radius.
The HCN emission is more confined to the inner regions than IR and CO.

The HCN/CO, IR/HCN (SFE$_{\rm dense}$) and IR/CO (SFE) distributions are shown in Figure 6. 
The HCN/CO and IR/CO ratios are much higher in the center than in the outer disk, but IR/HCN is lower. 
The IR/HCN ratio is $388 L_{\odot}(K\ km\ s^{-1}\ pc^2)^{-1}$ on average in the central 45\arcsec (1.66 kpc), while it is $691 L_{\odot}(K\ km\ s^{-1}\ pc^2)^{-1}$ in the outer disk, illustrating that the physical conditions in the center are quite different from the outer disk.

A pattern of high $L_{\rm IR}/L_{\rm CO}$ (SFE) and $L_{\rm IR}/L_{\rm HCN}$ (SFE$_{\rm dense}$) can be seen in some regions of the outer spiral arms (Figure 6). 
GMAs cores in the spiral arm are found to have nearby star-forming regions \citep{Egusa2011ApJ...726...85E} which suggests that star formation can be triggered in spiral arm.
\citet{Koda2012ApJ...761...41K} study CO lines in M51 and find that the CO(2-1)/CO(1-0) ratio is enhanced  in the spiral arms, especially downstream of the molecular arms. 
Higher resolution and sensitivity HCN observations are needed to study the effect of spiral arms on the HCN-IR correlation.

\subsection{Mass of Dense Molecular Gas Traced by HCN} %\label{bozomath}

As the measured IR--HCN correlation physically comes from the connection between star formation activity and dense molecular gas.
It should be noted that the mass of molecular gas estimated from HCN include considerable uncertainties because HCN emission is optically thick.
Sandstrom et al. (2013) find that in the HERACLES sample the CO emission to H$_2$ factor ($M_{\rm H_2}/L_{\rm CO}$) increases from the central kiloparsec to regions further out in the disk, in agreement with earlier work \citep[e.g.][]{Nakai1995PASJ...47..761N, Sodroski 1995ApJ...452..262S, Braine1997A&A...326..963B}. 
This suggests that $M_{\rm dense}/L_{\rm HCN}$ may increase with galactocentric radius as well.
In this situation,  the SFR-$M_{\rm dense}$ correlation may be linear while $L_{\rm IR}-L_{\rm HCN}$ may not, requiring a varying  $M_{\rm dense}/L_{\rm HCN}$ factor to convert from $L_{\rm HCN}$ to $M_{\rm dense}$.

\citet{Gao2004ApJS..152...63G} suggest that $M_{\rm dense}/L_{\rm HCN}$ may vary linearly with the gas excitation temperature.
They estimate that for an excitation temperature of 35 K,  $M_{\rm dense}/L_{\rm HCN} \sim 10 M_\odot (K\ km\ s^{-1}\ pc^2)^{-1}$.
The position-dependent $M_{dense}/L_{HCN}$ conversion can then be estimated as:  $M_{\rm dense}/L_{\rm HCN}  = 10 \times (35/T) M_\odot (K\ km\ s^{-1}\ pc^2)^{-1}$.
Since we have no means of estimating the excitation temperature, we assume that the dust temperature is representative of the temperature of the gas in dense regions.  This is probably not entirely unreasonable as the dust and gas exchange energy fairly quickly at high density.
The dust temperatures are estimated from the Herschel 160 $\micron$ and 250 $\micron$ fluxes  assuming an emissivity index of $\beta = 2$. 
While most spirals show a clear decrease in dust temperature with radius, in M51 we do not find a radii trend and the dust temperature varies little (24--28 K) in the central disk and outer spiral arm regions, resulting in little change in $M_{\rm dense}/L_{\rm HCN}$. 

There are still some weaknesses for our estimation of $M_{\rm dense}/L_{\rm HCN}$. The gas excitation temperature may be different from the dust temperature in GMC cores. And we do not consider the decreasing of HCN abundance which may be the primary cause of the apparent increase of $M_{\rm dense}/L_{\rm HCN}$ with radius. 
Another possible cause for the higher average IR/HCN ratio in the outer disk is that some of the IR emission may come from the general interstellar radiation field, which probably contributes proportionally more to the outer disk IR than to the center IR \citep{Calzetti 2005ApJ...633..871C}.

\subsection{Spatially-Resolved Star Formation Laws}

The spatially-resolved star formation laws quantify the relationship between the 􏰀SFR and molecular gas in nearby galaxies at scales of $\sim$ 1 kpc or less.
SFR--H$_2$ relation shows various power-law indexes, ranging from $< 1$ \citep[e.g.][]{Shetty2013MNRAS.430..288S}, $\sim 1$ (Bigiel et al. 2008), $\sim$ 1.1 -- 1.5 \citep{Kennicutt2007ApJ...671..333K, Schruba2010ApJ...722.1699S}.
A single linear spatially-resolved 􏰀SFR--H$_2$ relationship can not describe relation between star formation and molecular gas for the large scatter from galaxy to galaxy \citep{Saintonge2011MNRAS.415...61S, Schruba2011AJ....142...37S, Rahman2012ApJ...745..183R, Leroy2013AJ....146...19L}.

Correlations between SFR and dense molecular gas tracers, such as HCN 1--0 \citep{2015ApJ...805...31L}, HCN 4--3, HCO$^+$ 4--3, CS 7--6 \citep{Zhang2014ApJ...784L..31Z} and high-J CO \citep{Liu2015arXiv150405897L}, show a tight linear correlation in global galaxies. 
However, IR-HCN shows a sub-linear relation in M51 at kpc scale which is likely biased by the limited dynamical range, particularly the high  IR central regions.
And it is possible that the galaxy variations may dominate this inconsistency as shown in SFR--H$_2$ relation. 
More HCN survey toward galaxy is needed to explore the spatially-resolved IR--HCN relation.

\subsection{L$_{\rm HCN}$ vs. $L_{\rm IR}$: Correlation from GMCs to High-z Galaxies }

One of the motivations for studying HCN in M51 comes from the intimate link between star formation and dense gas in galaxies and in individual GMC cores in the Milky Way. 
In Figure 7, we show plots of $L_{\rm IR}$ versus $L_{\rm HCN}$ and $L_{\rm IR}$/$L_{\rm HCN}$ (an indicator of SFE$_{\rm dense}$) versus $L_{\rm IR}$ measured locally in M51 (center, filled black triangles; outer disk, open triangles) compared with the observations of  individual GMC cores in the Milky Way \citep[open squares;][]{Wu2010ApJS..188..313W}, M33 \citep[filled black squares;][]{Buchbender2013A&A...549A..17B}, M31 \citep[filled blue squares;][]{Brouillet2005A&A...429..153B}, nearby galactic centers \citep[filled blue triangles;][]{Krips2008ApJ...677..262K}, whole galaxies \citep[open hexagons;][]{Gao2004ApJS..152...63G} and high-z galaxies \citep[filled hexagons;][]{Gao2007ApJ...660L..93G}. 
For whole galaxies, the mean value of $L_{\rm IR}$/$L_{\rm HCN}$ is $900 L_{\odot}(K\ km\ s^{-1}\ pc^2)^{-1}$ (shown as the solid line), while it is $629 L_{\odot}(K\ km\ s^{-1}\ pc^2)^{-1}$ in M51. 
The $L_{\rm IR}$/$L_{\rm HCN}$ ratio in M51 is a factor 1.5 lower on average than that of most galaxies, yet is still contained in the scatter.
This suggests that a linear correlation is highly applicable to different physical scales, linking GMCs to kpc local regions and whole galaxies in a single relation.
The linearity of the HCN--IR correlation was interpreted as a fundamental unit of all star-forming system, thus both IR and HCN are simply accumulated by adding in more units \citep{Wu2005ApJ...635L.173W} . 

In M33, \citet{Buchbender2013A&A...549A..17B} observed 7 clouds at roughly 100pc resolution in CO, HCN and HCO$^+$. It is apparent from Figure 7 that the HCN emission for a given IR luminosity is weak in M33 (filled black squares), a factor 3 on average weaker than in the disk of M51.  
A reason to expect the HCN line to be weak with respect to IR is the sub-solar metallicity in M33. The dust (where IR from) abundance is expected to vary approximately linearly with metallicity (i.e. O/H) assuming a similar fraction of heavy elements in dust phase. Because N abundance decreases more quickly than the O abundance (i.e. N/O ratio decreases with decreasing O/H) \citep[see Figure 18 of review by ][]{Garnett2002astro.ph.11148G}, the HCN line is expected to decrease more than IR relative to Galactic abundances in M33. 
The disk of M31 was observed at a similar resolution by \citet{Brouillet2005A&A...429..153B}; using the IR luminosity from the \citet{Draine 2014ApJ...780..172D} maps, these points are also shown in Figure 7 (filled blue squares) and are in rough agreement with the M51 IR/HCN ratio.  From the resolved observations, the IR/HCN ratio is low for galactic centers, moderate (i.e. following the Gao et al. relation) for the disks of M51 and M31, and high for M33.  This whole range is covered by the individual galactic clouds studied by \citet{Wu2010ApJS..188..313W,Wu2005ApJ...635L.173W}.

The centers of 9 nearby galaxies NGC 1068, 2146, 3627, 4569, 4826, 5194, 6946, M82 and Arp 220 have been observed in HCN(J = 1 - 0) with the IRAM 30 m \citep{Krips2008ApJ...677..262K}.  These objects  exhibit nuclear star burst and/or AGN signatures. We estimate the infrared luminosity of the centers from the convolved Herschel 100 $\micron$ map \citep{Galametz2013MNRAS.431.1956G} observed by KINGFISH\footnote{Key Insights on Nearby Galaxies: a Far-Infrared Survey with Herschel} or the combination of the convolved Herschel 70 and 160 $\micron$ map (see section 3.2.1) observed by VNGS.
NGC 2146 and M82 centers show higher $L_{\rm IR}$/$L_{\rm HCN}$ ratios than M51, 
the other 7 nearby galactic centers are consistent with the central regions of M51.
The lower $L_{\rm IR}$/$L_{\rm HCN}$ ratios of nearby galactic centers compare to the mean value of whole galaxies (combination of outer disk and center ) suggest that the $L_{\rm IR}$/$L_{\rm HCN}$ ratio may be higher in the outer disk than the center, which support the increasing trend from center to outer disk in M51. More HCN observations toward the outer disk are needed to study this trend.

\begin{figure}
\centering\includegraphics[angle=0,scale=.7]{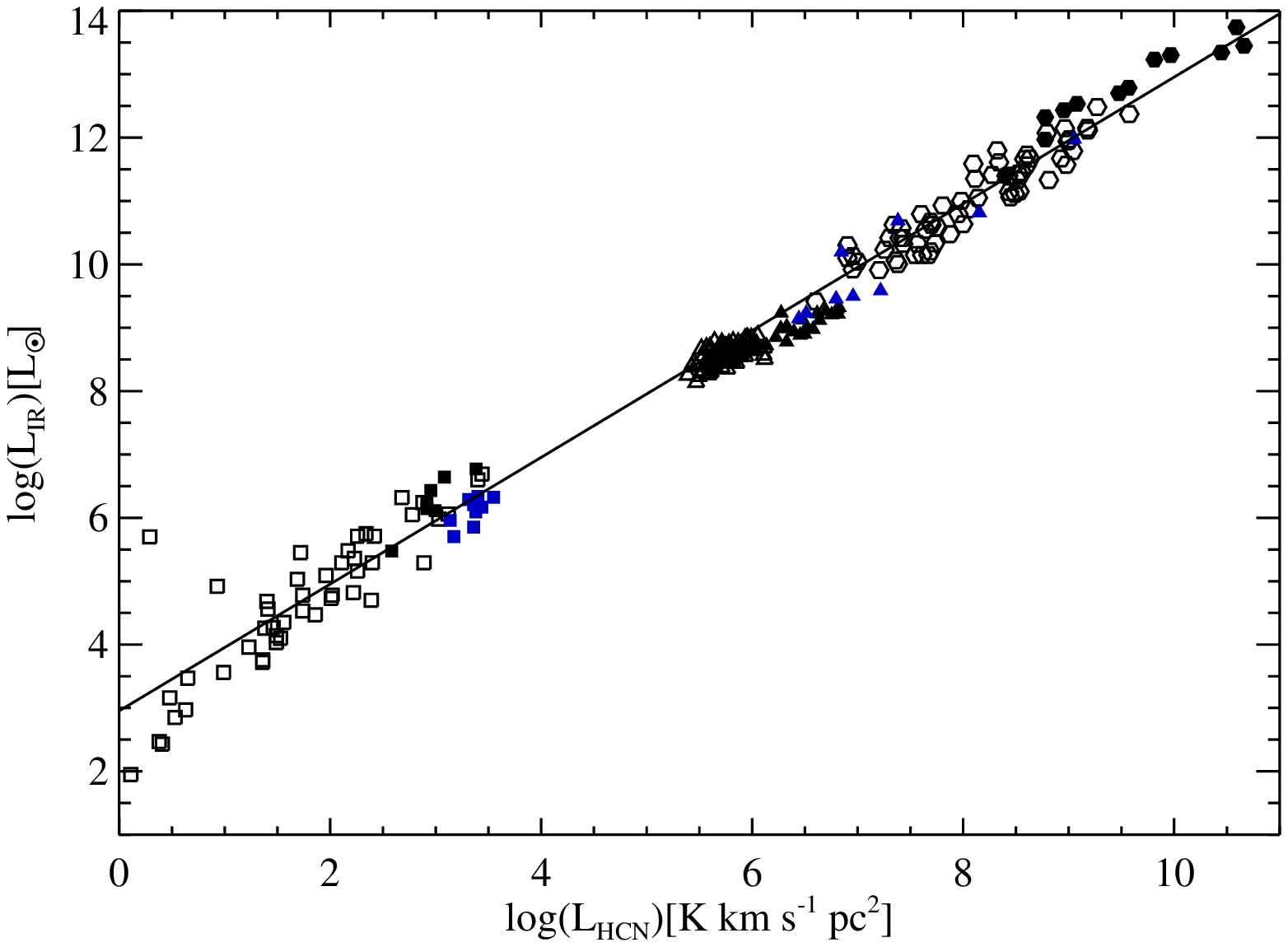}
\centering\includegraphics[angle=0,scale=.7]{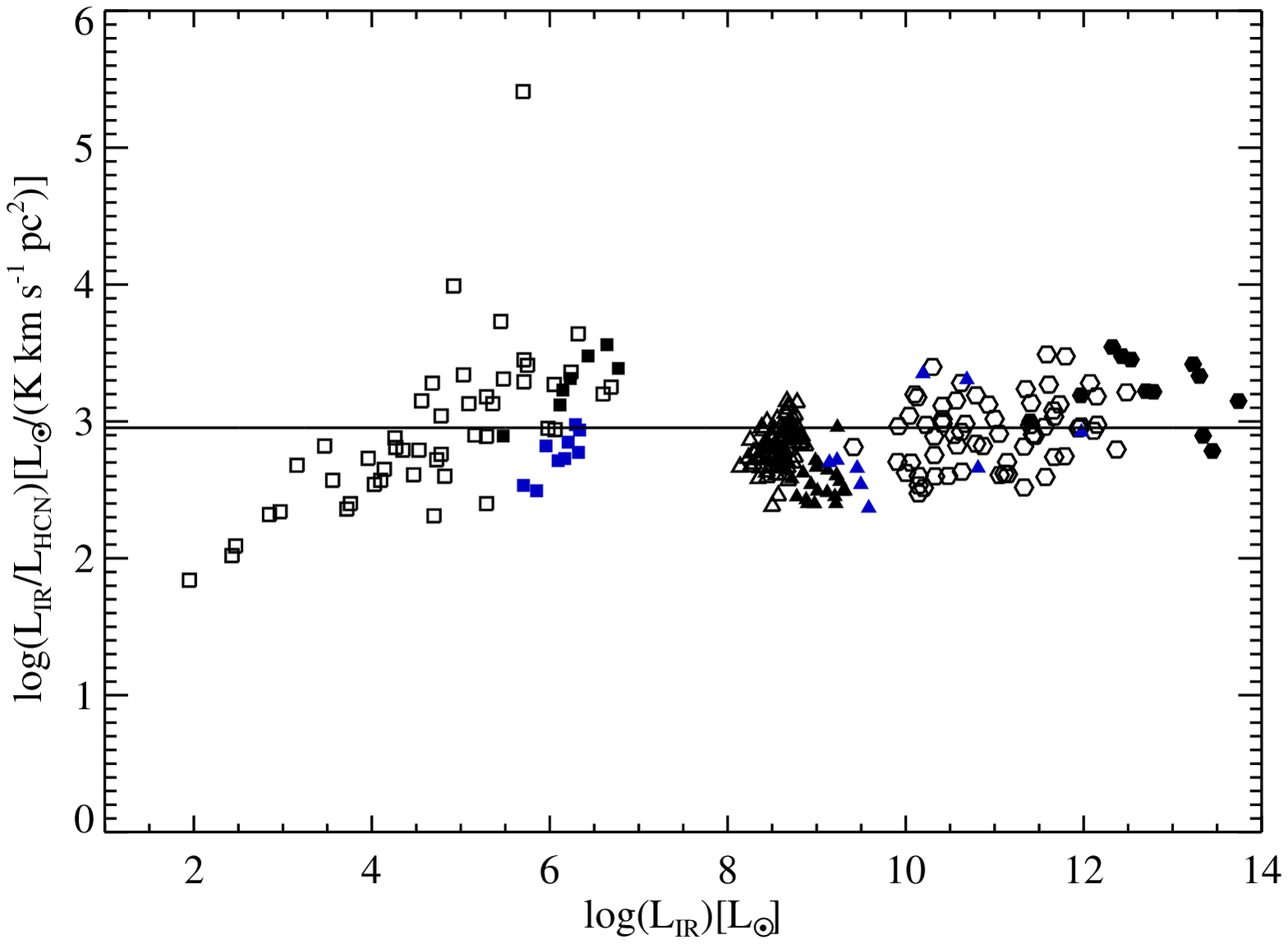}
\caption{ $L_{\rm IR}$-$L_{\rm HCN}$ correlation (upper) and $L_{\rm IR}$/$L_{\rm HCN}$-$L_{\rm IR}$ correlation (lower) for M51 observations (open and filled triangles), M33 (filled black squares), M31 (filled blue squares), nearby galactic centers (filled blue triangles), whole galaxies (open hexagons), high-z galaxies (filled hexagons) and cloud cores (open squares).} 
\end{figure}

\begin{figure}
\centering\includegraphics[angle=0,scale=0.9]{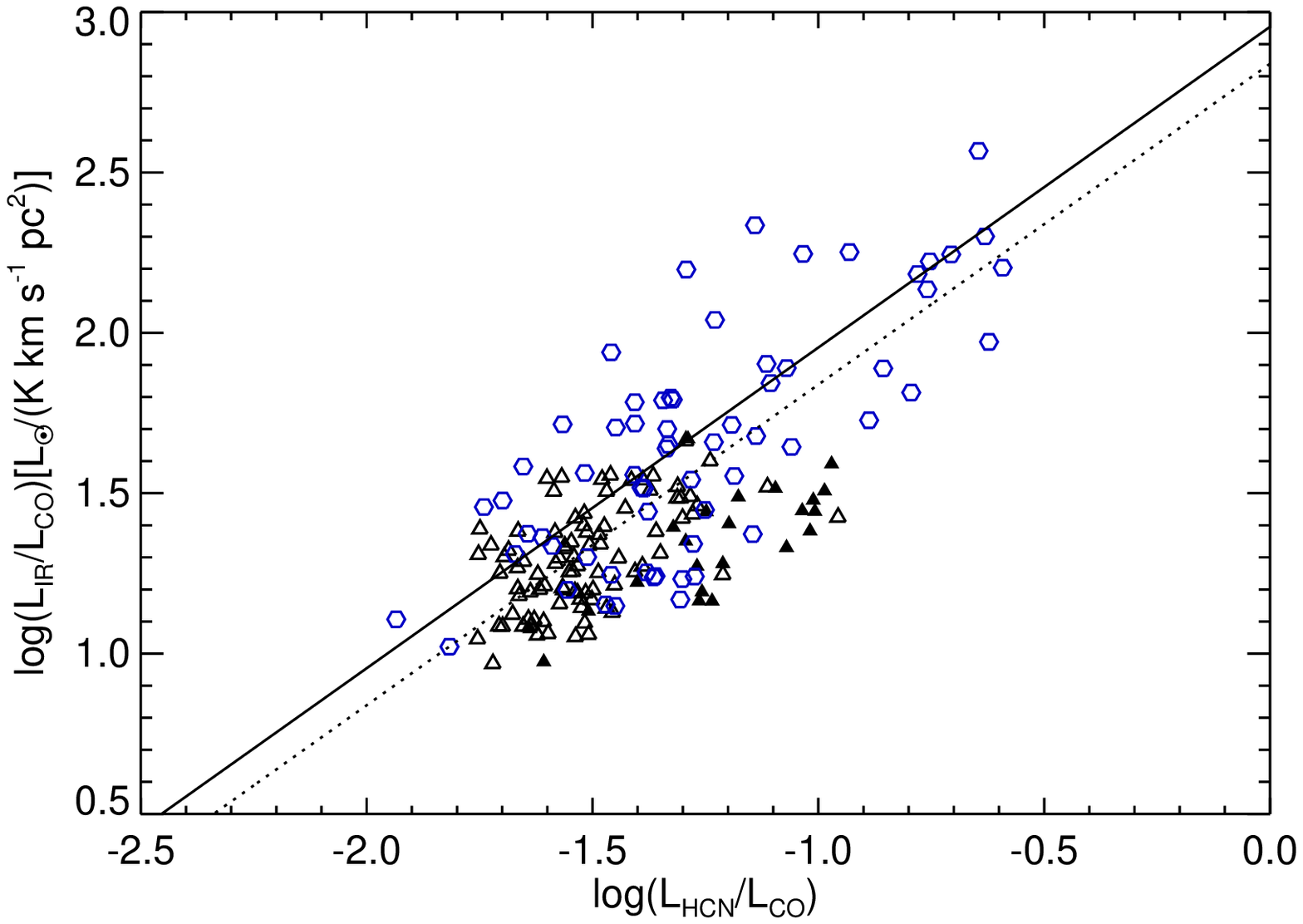}
\caption{Correlation between $L_{\rm IR}$/$L_{\rm CO}$ (trace SFE) and $L_{\rm HCN}$/$L_{\rm CO}$ ($R_{\rm HCN/CO}$) for the regions of M51 and global galaxies. Central (filled triangles) and outer disk (open triangles) regions in M51, galaxies (open blue hexagons) are distinguished in this figure. Solid line shows the correlation between $L_{\rm IR}$/$L_{\rm CO}$ and $L_{\rm HCN}$/$L_{\rm CO}$ for the global galaxies from \citet{Gao2004ApJ...606..271G} and dotted line show the consistent ratio of $L_{\rm IR}$/$L_{\rm CO}$ and $L_{\rm HCN}$/$L_{\rm CO}$ for the outer disk regions.}
\end{figure}

\subsection{$L_{\rm IR}$/$L_{\rm CO}$ vs. $L_{\rm HCN}$/$L_{\rm CO}$}

Figure 8 shows a correlation between $L_{\rm IR}$/$L_{\rm CO}$ (tracer of star formation efficiency; SFE) and $L_{\rm HCN}$/$L_{\rm CO}$ (tracer of dense gas fraction; $R_{\rm HCN/CO}$). 
The SFE in the central regions is systematically higher than the outer disk regions, which agrees well with \citet{Gao2004ApJ...606..271G} that the star formation efficiency depends on the fraction of molecular gas in a dense phase. However, the correlation between SFE and $R_{\rm HCN/CO}$ is not linear. $R_{\rm HCN/CO}$ in the center is higher than the linear fit determined from the mean value of the outer disk regions. 
Thus, either ($i$)  $L_{\rm IR}$ underestimates the SFR in the center (relative to the outer disk) or ($ii$) $L_{\rm HCN}$ overestimates $M_{\rm dense}$ in the center (relative to the outer disk) or ($iii$) the central environment makes star formation less efficient.
The 24 $\micron$ emission is less sensitive to the general interstellar radiate field and shows the same trends as in Figure 4, so we do not support ($i$). 
Our analysis excludes temperature as the primary cause of ($ii$). 
Abundance variations, presumably driven by Nitrogen, could provide at least a partial explanation for the trends we find.  Our work cannot address ($iii$).
To further study the effect of density on star formation, we are actually observing the higher dense gas tracer (HCN 4--3) in the central disk of M51 with the JCMT telescope, potentially extended the IR and HCN 4--3 relation in the global galaxies \citep{Zhang2014ApJ...784L..31Z}.

\section{ SUMMARY }

We present the spatially-resolved observations of the HCN J = 1 - 0 transition in the nearby spiral galaxy M51 using the IRAM 30 m telescope. The map covers an extent of $4\arcmin\times5\arcmin$ with spatial resolution of 28$\arcsec$ and 15$\arcsec$ sampling, which is, so far, the largest HCN map of M51. 
The HCN emission peaks 15$\arcsec$ west of the optical nucleus and about 15$\arcsec$  north of the CO peak (at the HCN resolution).  

Comparing the individual observed positions, the IR emission varies less than the HCN emission, such that $L_{\rm IR} \propto L_{\rm HCN}^{0.74\pm0.16}$, despite the overall linear relation between HCN and IR emission from Galactic clumps to distant galaxies. 
While the conversion from HCN intensity to dense gas mass may vary, temperature (as traced by the dust) cannot explain the variation.
It is possible that the galaxy variations may dominate this inconsistency as shown in SFR–-H$_2$ relation.

$L_{\rm IR}$ traces SFR and $L_{\rm HCN}$ traces dense gas mass, so the ratio provides information about the efficiency of transformation from dense gas into stars.
There is a clear radial trend in $L_{\rm IR}$/$L_{\rm HCN}$ decreasing with galactocentric radius.
Comparing with resolved data in other galaxies,
7 nearby galactic centers show a consistent $L_{\rm IR}$/$L_{\rm HCN}$ ratio (an indicator of SFE$_{\rm dense}$) with the central regions in M51. While the ratio measured locally in M51 is 1.5 times lower  than the galactic average, 
it is still within the scatter.
Though  $L_{\rm IR}$/$L_{\rm HCN}$ may vary with physical environments in M51, HCN and IR indeed shows a linear correlation over 10 orders of magnitude.

\acknowledgments

We are very grateful to Jin Koda for providing the CO 1--0 data on NGC 5194. The VNGS data was accessed through the Herschel Database in Marseille (HeDaM\footnote{http://hedam.lam.fr}) operated by CeSAM and hosted by the Laboratoire d'Astrophysique de Marseille. We appreciate the generous support from IRAM staff during the observations and data reduction. This work is partially supported by NSFC grant Nos. 11173059, 11390373, 11273015 and 11133001), CAS No. XDB09000000 and the National Basic Research Program (973 program No. 2013CB834905).

\end{document}